\documentstyle[12pt,epsf]{article}

\large
\newcommand{\be}{\begin{eqnarray}}
\newcommand{\ee}{\end{eqnarray}}
\newcommand\del{\partial}

\begin{document}
\setlength{\baselineskip}{21pt}
\pagestyle{empty}
\vfill
\eject
\begin{flushright}
SUNY-NTG-95/32
\end{flushright}

\vskip 2.0cm
\centerline{\Large  Universal scaling of the valence quark mass dependence}
\vskip 1cm
\centerline{\Large of the chiral condensate}
\vskip 2.0 cm
\centerline{\bf  J.J.M. Verbaarschot}
\vskip .2cm
\centerline{Department of Physics}
\centerline{SUNY, Stony Brook, New York 11794}
\vskip 2cm

\centerline{\bf Abstract}
Recently, the Columbia group obtained the valence quark mass dependence
of the chiral condensate for lattice QCD simulations with dynamical
fermions for a series of couplings close to the critical temperature.
In this letter we show that the data for temperatures below $T_c$
and sufficiently small valence quark masses can be rescaled to fall on a
universal curve. Using chiral random matrix
theory we obtain an exact analytical expression for the shape of this curve.
We also discuss the universal scaling behavior of the mass dependence
of the mass derivatives of the chiral condensate (scalar susceptibilities).

\vfill
\noindent
\begin{flushleft}
May 1995
\end{flushleft}
\eject
\pagestyle{plain}

Numerical simulations of QCD have contributed greatly
to our present understanding of the QCD chiral phase transition
(see \cite{DeTar} for a recent review).
It is now generally believed that the critical temperature is about
150 MeV, and that, for two light flavors, the phase transition is
of second order \cite{Brown}. In the neighborhood of the critical temperature
the particle spectrum changes drastically \cite{DeTar-Kogut-Karsch}
which may lead to important experimental implications.
The hadron masses are closely related to the chiral
condensate, the order parameter of the chiral phase transition, which,
according to the Banks-Casher formula \cite{BANKS-CASHER-1980},
is proportional to
the average spectral density of the Dirac operator in the vicinity of zero.

Unfortunately, the volumes of present day lattice calculations are
relatively small. On the other hand, the smallest eigenvalues of
the lattice QCD Dirac operator show the most significant volume dependence.
In order to extract the chiral condensate
reliably from lattice QCD simulations detailed knowledge of the volume
dependence of the spectral density near zero is essential.
Recent lattice calculations of the Columbia group \cite{Christ}
have probed this part
of the spectrum via the so called valence quark mass dependence of the
chiral condensate. In this paper we attempt to understand
this dependence for masses well below the typical hadronic scale $\Lambda$.
We also give results for the derivative of the chiral condensate
(scalar susceptibilities) in this mass region.

The valence quark mass dependence of the chiral condensate is defined by
\be
\Sigma^V(m) = \frac 1N \int_0^\infty \frac{2m\rho(\lambda)
d\lambda}{\lambda^2+m^2},
\ee
where $\rho(\lambda)$ is the average eigenvalue density of the Dirac operator
\be
\rho(\lambda) = \sum_n \langle \delta(\lambda -\lambda_n)\rangle,
\ee
and $N$ is the total number of eigenvalues
of the lattice QCD Dirac operator. The average over gauge field
configurations, denoted by $\langle \cdots \rangle$, is weighted by
the gluonic action and the fermion determinant with sea quark mass equal to
$m_{\rm sea}$. It is clear that, for a finite number of
eigenvalues, $\Sigma^V(m) \sim 1/m$ for asymptotically large valence quark
masses. For valence quark masses well below the average position
of the smallest eigenvalue, $\lambda_{\rm min}$, we
expect that $\Sigma^V(m) \sim m$. For masses larger than $\lambda_{\rm min}$
but still sufficiently small we hope to see a plateau which can be
identified as the chiral condensate (this assumes a smooth sea quark
mass dependence of the chiral condensate). The usual relation between
the chiral condensate $\Sigma$ and the average spectral density is given
by the Banks-Casher \cite{BANKS-CASHER-1980} formula
\be
\Sigma = \frac{\pi \rho(0)}{N}.
\ee
Here, $\rho(0)$ is defined as the extrapolation to zero
of the eigenvalue density many spacings away from zero.
This can be done unambiguously
if there are many eigenvalues below a typical hadronic scale $\Lambda$
(a more rigorous definition is obtained by taking the limits
${m\rightarrow 0}$ and ${N\rightarrow\infty}$ in this order).
Above this scale the eigenvalue density is dominated by the perturbative
eigenvalue density $\sim N\lambda^3$. In lattice calculations the latter
dependence has to be adjusted for finite size effects.
Because of the existence of a finite chiral condensate the Banks-Casher
formula tells us that the smallest eigenvalues of the Dirac operator
are spaced as $\sim \pi/N\Sigma$. This allows us to define a different limiting
procedure of the eigenvalue density of the Dirac operator, namely, the
microscopic spectral density defined by \cite{SV-1993}
\be
\rho_S(z) = \lim_{N\rightarrow \infty} \frac 1{N\Sigma}\rho(\frac z{N\Sigma}).
\ee
In terms of this density, eq. (1) can be rewritten as
\be
\Sigma^V(m) = \int_0^\infty \frac{2mN\Sigma \rho_S(z)dz}{z^2 + m^2N^2
\Sigma^2},
\ee
which suggests to plot the lattice data as $\Sigma(m)/\Sigma$ versus
$mN\Sigma$.

In Fig. 1 the points show the original Columbia data \cite{Christ} obtained
from a $16^3\times 4$ lattice with dynamical (staggered) fermions ($N_f = 2$)
with mass $m_{\rm sea}a= 0.01$ ($a$ is the lattice spacing).
Results are given for three different
values of $\beta$ (see label of the figure) below $T_c$.
The smallest eigenvalue is of
order $2.5\,10^{-4}$ for $\beta= 5.245$ and increases gradually
for larger values of $\beta$.
A clear plateau is observed for two
smallest values of $\beta$. It is located
around $ma \approx 10^{-3}$.
The dotted curves are fits to the lattice data for masses larger than
$ma\approx
0.001$. However, we did not fit directly the valence quark mass dependence
but used a reasonable fitting function for  the eigenvalue density, which
allowed us to extract $\rho(0)/N$.
\newpage
\vskip 0.5 cm

\noindent
\begin{figure}[t]
\begin{center}
\leavevmode
\epsffile{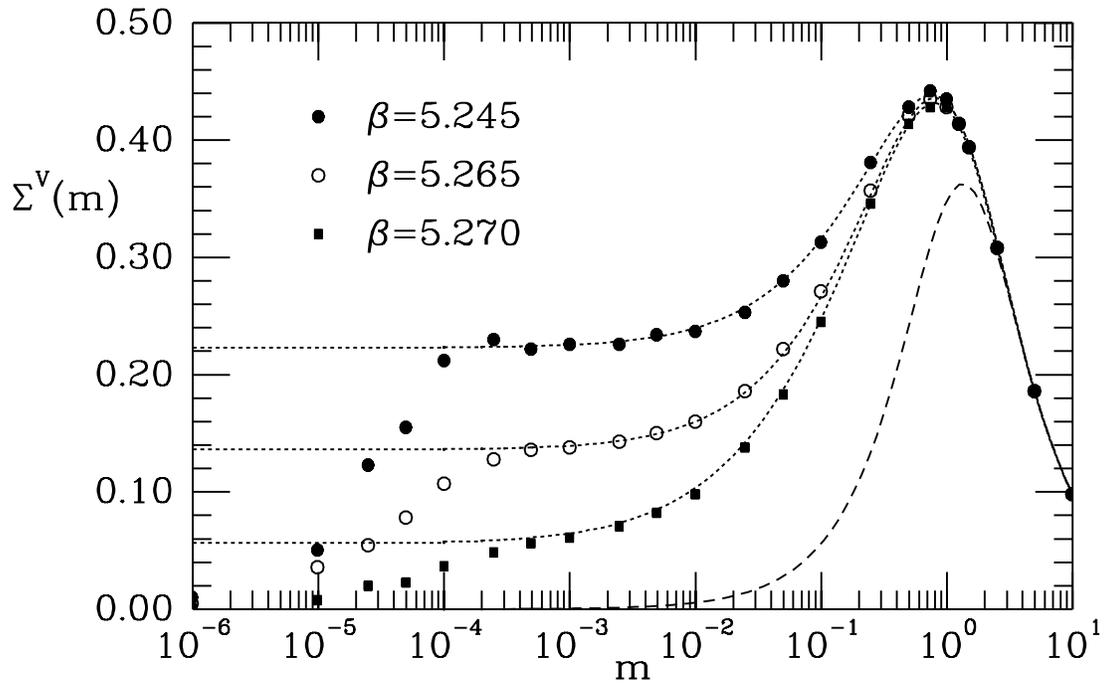}
\end{center}
\oddsidemargin3.7cm
\evensidemargin3.7cm
 \setlength{\textwidth}{16cm}
\caption{The valence quark mass dependence of the chiral condensate.
The points represent the Columbia data for values of $\beta$ as indicated
in the label of the figure. Fits to their data are shown by the dotted curves.
The result for the noninteracting Dirac operator is given by the dashed curve.}
\oddsidemargin1.7cm
\evensidemargin1.7cm
 \setlength{\textwidth}{18cm}
\end{figure}

In Fig. 2 we show the rescaled data for the same values of $\beta$. Remarkably,
they fall on a single curve suggesting that the microscopic
spectral density is a universal function.
Below  we obtain an analytical expression for the shape of this curve
(shown by the full line for $N_f= 0$ in this figure).

{}From the analysis of spectra of complex systems we know that the
correlations between eigenvalues of the Hamiltonian on a scale of no
more than a finite number of level spacings can be described by a random
matrix theory with the appropriate anti-unitary (time reversal) symmetry
\cite{Bohigas,Mehta,Porter}.
The situation in the case of the Dirac operator is more complicated.
First of
all, because of the $U_A(1)$-symmetry, in a chiral basis the
massless Dirac operator has the block structure (for zero quark masses)
\be
\left( \begin{array}{cc} 0 & T\\ T^\dagger &0 \end{array}\right).
\ee
Since both the gauge fields and the Dirac matrices are in general complex,
the anti-unitary symmetries act both in color and Dirac space. For a detailed
discussion of this classification we refer to \cite{V}. At this moment
we only want to remark that for three colors with fundamental fermions
there are no anti-unitary symmetries, leading to complex
matrix elements $T_{ij}$. This ensemble was called the chiral
gaussian unitary ensemble (chGUE).

\vskip 0.5 cm
\noindent
\begin{figure}[t]
\begin{center}
\leavevmode
\epsffile{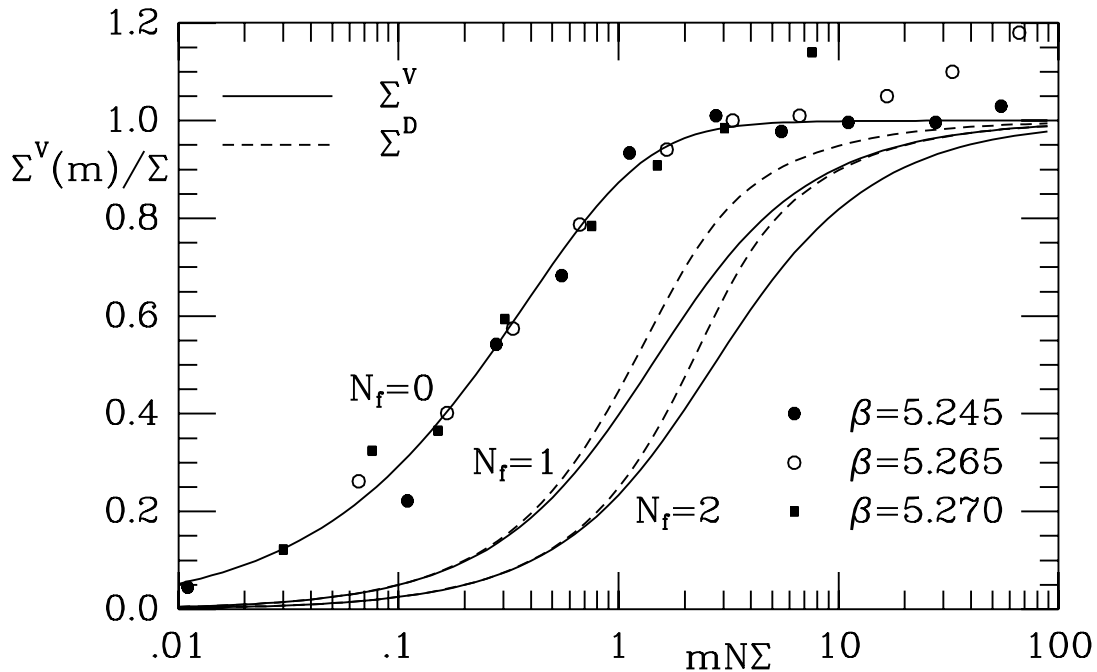}
\end{center}
\oddsidemargin3.7cm
\evensidemargin3.7cm
 \setlength{\textwidth}{16cm}
\caption{The valence quark mass dependence of the chiral condensate
$\Sigma^V(m)$ plotted as $\Sigma^V(m)/\Sigma$ versus $m N\Sigma$. The dots and
squares represent lattice results obtained by the
Columbia group [1] for
values of $\beta$ as indicated in the label of the figure. The full lines
correspond to the random matrix result (eq. (8)) for $\nu =0 $ and $N_f$
as indicated in the label of the figure. Also shown are results (dashed curves)
for the quark mass dependence with equal valence and sea quark masses.}
\oddsidemargin1.7cm
\evensidemargin1.7cm
 \setlength{\textwidth}{18cm}
\end{figure}

Because of the chiral structure of the Dirac operator, all nonzero
eigenvalues occur in pairs $\pm \lambda_n$. Therefore, $\lambda= 0$ is a
special point in the spectrum. This allows us  consider two types of level
correlations: i) microscopic correlations in the bulk of the spectrum;
they are given by the Dyson-Mehta-Wigner random matrix ensembles and ii)
the microscopic spectral density (defined in (4)).
At this moment it is not clear in
how far deformations of the random matrix ensemble while keeping its
chiral structure (6) lead to the destruction of both types of universal
correlations at the same time,  or that one of them is more strongly universal
than the other. Up to know we have not been able to construct a
random matrix ensemble for which the latter is true \cite{Horner}.
This is an important
point because recently Kalkreuter \cite{Kalkreuter} obtained all eigenvalues
of the Dirac operator for a $12^4$ lattice with dynamical (staggered) fermions.
We have analyzed the eigenvalue correlations in the bulk of the spectrum
and found \cite{HV} that they are given by the Dyson-Mehta-Wigner
random matrix ensembles. If the two types of universal behavior are
mutually inclusive, which we believe to be case, the microscopic spectral
density of the Dirac operator is given by random matrix theory as well.
Further arguments in favor of this conjecture can be found in \cite{Vmin}.
General universality arguments \cite{BZ}
show that the properties of random matrix
ensembles are not sensitive to specific shape of the distribution of
the matrix elements. For convenience, and in agreement with the maximum
entropy principle \cite{Balian}, we have chosen a gaussian distribution.

The microscopic spectral density of the corresponding chiral Gaussian
Unitary Ensemble was obtained in \cite{VZ,V} and is given by
\be
\rho_S(z) = \frac z2(J^2_{N_f+\nu}(z) -J_{N_f+\nu+1}(z)J_{N_f+\nu-1}(z)).
\ee
Here, $N_f$ is the number of (massless) flavors and $\nu$ is the topological
charge. Using (5) we are able to calculate the valence quark mass dependence
of the chiral condensate. The integrals are known analytically
\cite{Graedshteyn} which leads to a remarkably simple result
\be
\frac{\Sigma^V(u)}{\Sigma} = u (I_{N_f+\nu}(u) K_{N_f +\nu}(u) +
I_{N_f+\nu+1}(u) K_{N_f +\nu-1}(u)),
\ee
where the $I_\nu$ and the $K_\nu$ are modified Bessel functions and
\be
u \equiv mN\Sigma.
\ee
The valence quark mass dependence for zero, one and two massless flavors
and topological charge $\nu = 0$ as given by eq. (8) is represented
by the full lines
in Fig. 2. Obviously, the lattice data are described by the curve
with $N_f = \nu = 0$. In hindsight this is no surprise. First,
in the region where the
valence quark mass is much less than
the sea quark mass, the fermion determinant has no bearing on the Dirac
spectrum and we have effectively $N_f =0$.
Second, on a lattice the zero mode states and the much larger
number of nonzero modes states are mixed, so that
the effective topological charge that
reproduces the lattice results is equal to zero.
The small $u$ behavior of the valence quark mass dependence is $-u\log u$ for
$N_f + \nu = 0$ and  linear for $N_f + \nu \ne 0$.
For valence quark masses smaller than the smallest eigenvalue of the Dirac
operator over the
$finite$ ensemble of QCD gauge field configurations,
a linear mass dependence is expected,
which is indeed observed in lattice gauge calculations.

For comparison we have also shown the sea quark mass dependence of
the chiral condensate (dashed line in Fig. 2). This can be obtained
most conveniently from the finite volume partition function which
follows from general symmetry arguments \cite{LS}
or can be derived starting from the chiral random matrix ensemble
\cite{Halasz}. The effective partition function describes the mass
dependence of the QCD partition function for space time volumes
\be
 \frac 1{\Lambda} \ll V^{1/4} \ll \frac 1{m_\pi}
\ee
($\Lambda$ is a typical hadronic scale and $m_\pi\sim m\Lambda$
is the pion mass). It is straightforward to obtain the sea quark mass
dependence of the chiral condensate from the effective partition function.
In the sector of zero topological charge one finds \cite{LS}
\be
\frac{\Sigma^S(u)}{\Sigma} = \frac{I_1(u)}{I_0(u)}\quad {\rm for}\quad
N_f = 1,
\ee
and
\be
\Sigma^S(u) = \frac{I_1^2(u)}{u(I_0^2(u)-I_1^2(u))}\quad {\rm for}\quad
N_f = 2.
\ee
These results
are shown by the dashed curves in Fig. 2 and should be compared to
lattice data for
the $sea$ quark mass dependence of the chiral condensate which,
unfortunately, are not available at this moment.
For $m \ll \lambda_{\rm min}$
the mass in the determinant in not relevant and $\Sigma^c(u)$ should
approach $\Sigma^V(u)$ which is indeed seen in Fig. 2.

For what range of masses can we expect the valence quark mass dependence to
be described by random matrix theory? For $N_f \ne 0$,
the small mass limit chiral random matrix theory reduces
to the partition function of Leutwyler and Smilga \cite{Halasz}, which gives
the $m$-dependence of the QCD partition function in the range (10). The
$N_f =0$ limit can be obtained from results at arbitrary $N_f$ by
taking the limit $N_f \rightarrow 0$, in analogy with the replica trick in
condensed matter physics\footnote{Alternative, one may use a supersymmetric
formulation to cancel the fermion determinant leading to supersymmetric
quenched chiral perturbation theory \cite{Golterman}.}.
Since the condition (10) is $N_f$ independent it should
be valid for $N_f = 0$ as well
(with $m_{\pi} = m \Lambda$ and $m$ equal to the valence quark mass).
On the other hand, for lattice calculations to be physical,
the pion should fit  well inside the box,
\be
\frac 1{\sqrt{m_{\rm sea} \Lambda}} \ll L,
\ee
which leads to the condition
\be
m \ll \frac 1{L^2\Lambda} \ll m_{\rm sea}.
\ee
Because the pion is the lightest hadron the condition $\frac 1/\Lambda \ll L$
in (10) is satisfied automatically. In practice the pion compton wave length
is not $much$ smaller than the length of the box. Indeed, the lattice
data show a universal behavior up the valence quark masses close to
the sea quark mass.

It is natural to expect that also the derivatives of the chiral
condensate with respect to the mass (the scalar susceptibilities)
show a universal scaling. Below we will discuss the following
susceptibilities
\be
\chi^{\rm dis} &=& \frac 1N \del_{m_f}\del_{m_{f'}} \log Z
=\langle\frac 1N \sum_{k,l=1}^N \frac 1{i\lambda_k + m}
\frac 1{i\lambda_l + m}\rangle -
N\langle  \frac 1N \sum_{k=1}^N \frac 1{\lambda_k^2 +m^2}\rangle^2,\\
\chi^{\rm con} &=& \frac 1N \del_{m_f} \del_{m_f} \log Z -\chi^{\rm dis}
=-\langle \frac 1N \sum_{k=1}^N \frac 1{(i\lambda_k+m)^2}
\rangle.
\ee
where the quark masses have been put equal to $m$ after differentiation. The
latter quantity is the isovector scalar susceptibility and $\chi^{\rm dis}$
enters in the isoscalar scalar susceptibility, $\chi^{\sigma} =
2\chi^{\rm dis} + \chi^{\rm con}$ (for two flavors).


As follows immediately from (16) $\chi^{\rm con}$ can be calculated from
average spectral density (2). On the other hand,
$\chi^{\rm dis}$ can be expressed into
the connected two-point level correlation function
\be
\rho_c(\lambda,\lambda') = \langle \rho(\lambda)\rho(\lambda') \rangle
-\langle \rho(\lambda)\rangle \langle \rho(\lambda') \rangle.
\ee
If the eigenvalues are uncorrelated we
have
\be
\rho_c(\lambda,\lambda') = \langle
\rho(\lambda)\rangle\delta(\lambda-\lambda'),
\ee
which leads to
\be
\chi^{\rm dis} = \frac {\Sigma}{m}.
\ee

In random matrix theory, the eigenvalues are strongly correlated, which gives
rise to a quite different prediction.
We will consider the susceptibilities in the limit $mN\Sigma \gg 1$, but
with $m \ll \Lambda$. If, at the same time, $m^2 N \Lambda^2 \ll 1$, the
partition function of Leutwyler and Smilga or chiral random
matrix theory can be used to calculate the susceptibilities
\cite{HL}.
In this limit
the leading order contribution is expected to be of the form $1/m^2N$.
In the physical limit, where the $m^2 N \Lambda^2$ correction to the
vacuum energy cannot be neglected,
general symmetry arguments are insufficient to determine the susceptibility.

Since we are interested in the valence quark mass of the susceptibility we
cannot use the effective partition function but have to rely on chiral
random matrix theory. The valence quark mass dependence of $\chi^{\rm con}$
can be obtained from $\Sigma^V(u)$ given in eq. (8),
\be
 \chi^{\rm con} = \del_m \Sigma^V(u)\sim\frac{7}{32} \frac 1{m^2V} \frac 1u,
\ee
where the asymptotic result is valid for $u\gg 1$ (the coefficient
of the leading order term in $1u$ vanishes).
For $N_f =0$,
 $\chi^{\rm dis}$ can be obtained from the two level correlation function
\cite{VZ}. The result is given by
\be
\chi^{\rm dis} &=& \frac 1{4m^2 V}F^{\rm dis} (u),
\ee
where both $F^{\rm dis} (u)$
can be expressed in modified
Bessel functions and approaches 1 for $u\gg 1$.

In the limit $m^2 N\Lambda^2 \gg 1$ the susceptibility is
no longer determined by the lowest order terms in a chiral expansion
of the vacuum energy.
However, it is possible to calculate
the susceptibilities in the framework  of chiral random matrix theory
but the results are model dependent.
$\chi^{\rm con}$ can be obtained directly from
the semicircular eigenvalue distribution. $\chi^{\rm dis}$ could
only be obtained in an indirect way from the bosonized random matrix
partition function. The results for all $N_f$ are given by
(for $m^2 N\Lambda\gg 1)$
\be
\chi^{\rm con} &=& -\frac{\Sigma^2}2,\\
\chi^{\rm dis} &=& 0.
\ee
The result for $\chi^{\rm dis}$ is a consequence of the stiffness of the
random matrix spectra.

Unfortunately, no lattice data are available for the valence quark mass
dependence of the susceptibilities quoted above.
What has been calculated \cite{Karsch}
is the dynamical quark mass dependence
of the susceptibilities. Because, in this way only relatively large quark
masses have been studied, ultraviolet divergent terms make direct
comparison difficult. In $\cite{Karsch}$ $m^2V \approx 1$, but the numerical
results show that, below the critical temperature, the quark mass dependence
of the susceptibilities is $\sim 1/m$. One explanation is that
the eigenvalues of the Dirac operator are only weakly correlated
so that eq. (20) is applicable. On the other hand,
spectral averaging shows strong correlations between the eigenvalues
\cite{HV} implying that the
ensemble and the spectral average of the eigenvalue correlations are not
equal. This exciting possibility deserves further attention.

In conclusion, we have shown that the valence quark mass dependence of
the chiral condensate is lattice QCD is given by a universal curve that
can be described by chiral random matrix theory. As soon as lattice QCD
results for the valence quark mass susceptibilities are available,
we hope to extend our analysis to these observables as well.

\vglue 0.6cm
{\bf \noindent  Acknowledgements \hfil}
\vglue 0.4cm
 The reported work was partially supported by the US DOE grant
DE-FG-88ER40388. A. Halasz, A. Jackson and A. Smilga are acknowledged
for useful discussions.

\vfill
\eject
\newpage
\setlength{\baselineskip}{17pt}

\vfill
\eject

\end{document}